\documentstyle[11pt,ysc,twoside,epsf]{article}
\def\edcomment#1{\iffalse\marginpar{\raggedright\sl#1\/}\else\relax\fi}
\reversemarginpar

\begin{document}
\title{Period Change of Eclipsing Binaries from the ASAS Catalog}

\author{Radoslaw Poleski}
\affil{{\tt rpoleski@astrouw.edu.pl}, Warsaw University Astronomical
Observatory, Al. Ujazdowskie 4, Warsaw 00-478, Poland}

\author{Bogumil Pilecki}
\affil{{\tt pilecki@astrouw.edu.pl}, Warsaw University Astronomical
Observatory, Al. Ujazdowskie 4, Warsaw 00-478, Poland}

\begin{abstract}
We present a preliminary statistical analysis of a period change of
eclipsing binaries from the ASAS Catalog of Variable Stars. For each
contact and semidetached system brighter than 13.3$mag$ (in V) with
a period shorter than 0.4 days and at least 300 observation points
we have found an angular velocity $\omega$ and its time derivative
$\frac{d\omega}{dt}$. According to our accuracy there is no evidence
that average $\frac{d\omega}{dt}$ differs from 0. Light curves for
selected stars are presented.

\end{abstract}

\section{Introduction}

Eclipsing binaries are very important in astrophysical research
giving us much information not only about stars they are formed of
but also about stars in general. We can determine their dimensions
and masses as well as surface temperature and other physical
parameters (Wilson 1994). In particular, rate and sign of a period
change give us information about physical processes that occur in
these systems. Angular momentum loss (AML) by magnetic breaking
(van't Veer 1979; Rahunen 1981; Vilhu 1982 and others) may cause a
period decrease, mass transfer between components in thermal
relaxation oscillations (TRO) (Flannery 1976; Lucy 1976; Robertson
\& Eggleton 1977) may cause both increase and decrease of a period
and an influence of the third component may cause periodic
oscillations of its value.

Statistical analysis of period change rate can tell us more about
structure and evolution of binary systems. Here we present first,
preliminary results of this analysis considering short period
contact and semidetached binaries from ASAS-3 data.

\section{About ASAS-3}

The All Sky Automated Survey-3 is situated in the Las Campanas
Observatory and is operating since August 2000. The project main
goal is to monitor all stars brighter than 14$mag$. It consists of
two wide-field 200/2.8 instruments, one narrow-field 750/3.3
telescope and one super-wide 50/4 scope, each equipped with the CCD
camera. Observations are carried out with VRI filters.

Nowadays there are more then 2,300,000,000 photometric measurements for more
then 15,000,000 stars south of declination +28\deg. Among these stars more than 50,000 are
confirmed to be variables and most of them are new discoveries. Each field is observed once each
2 or 3 nights. The ASAS-3 Catalog of Variable Stars is presented on-line on the Internet
\footnote{\tt www.astrouw.edu.pl/\~{}gp/asas/asas.html} with light curves updated just after exposure. Full
description of the project was given by Pojmanski (2002).

\section{Data}

This first attempt to search for changing period eclipsing binaries
considered stars brighter than $13.3 mag$ (in V) with period $P <
0.4 d$. To assure better reliability for the results we further
restricted these stars to have at least 300 observation points.
These criteria limited the number of stars from about 8300 to 1099
contact (EC) and semidetached (ESD) stars. On average we had 420
observations per star. We have used only V filter data presented by
Paczynski et al. (2006).

For each star we have used a standard AOV method
(Schwarzenberg-Czerny 1989) to simultaneously determine an angular
velocity $\omega$ and its time derivative $\frac{d\omega}{dt}$. In
this method two dimensional power spectrum was created and then
model with maximum power was selected as the best fit to the data.

\section{Results}

\begin{figure}[ht]
\plotone{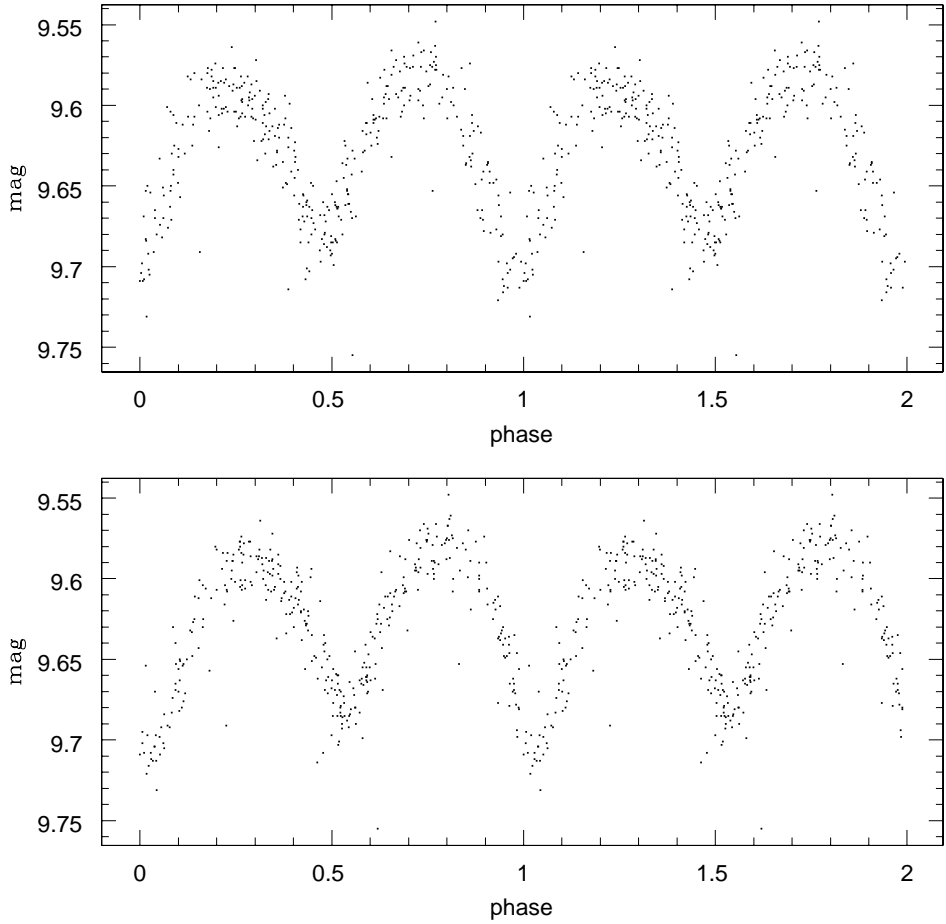}
\caption{Light curve of 004430-3606.5 star folded with constant (top) and varying (bottom) period.}
\end{figure}
\begin{figure}[ht]
\plotone{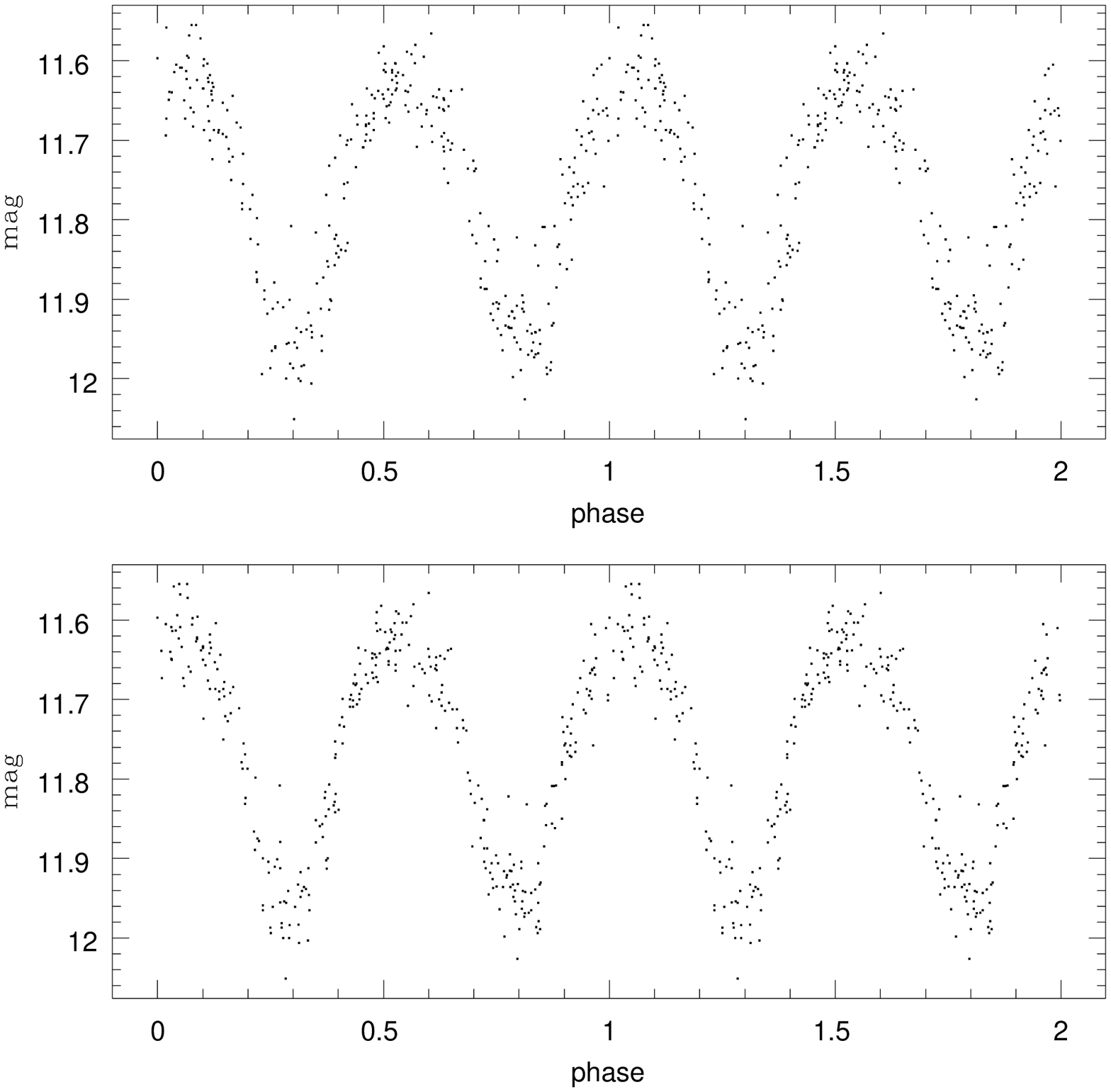}
\caption{Light curve of 082456-4833.6 star folded with constant (top) and varying (bottom) period.}
\end{figure}

Light curves for two selected stars are presented in Figures 1 and 2. It is easily seen
that considering a period change improves the shape of the light curve. For these stars
following parameters were found ($A$ --- power):\\

\noindent star 004430-3606.5:\\
$\bullet$ constant period \\
$\omega = 25.4857 d^{-1}$\\
$A = 2.410$\\
$\bullet$ with period change\\
$\omega = 25.4867 d^{-1}$\\
$\frac{d\omega}{dt}=-9.90\cdot10^{-7} d^{-2}$\\
$A=2.784$\\

\noindent star 082456-4833.6:\\
$\bullet$ constant period\\
$\omega = 17.22004 d^{-1}$\\
$A = 2.953$\\
$\bullet$ with period change\\
$\omega=17.21960 d^{-1}$\\
$\frac{d\omega}{dt}=-5.47\cdot10^{-7} d^{-2}$\\
$A=3.599$\\

\begin{figure}[ht]
\plotone{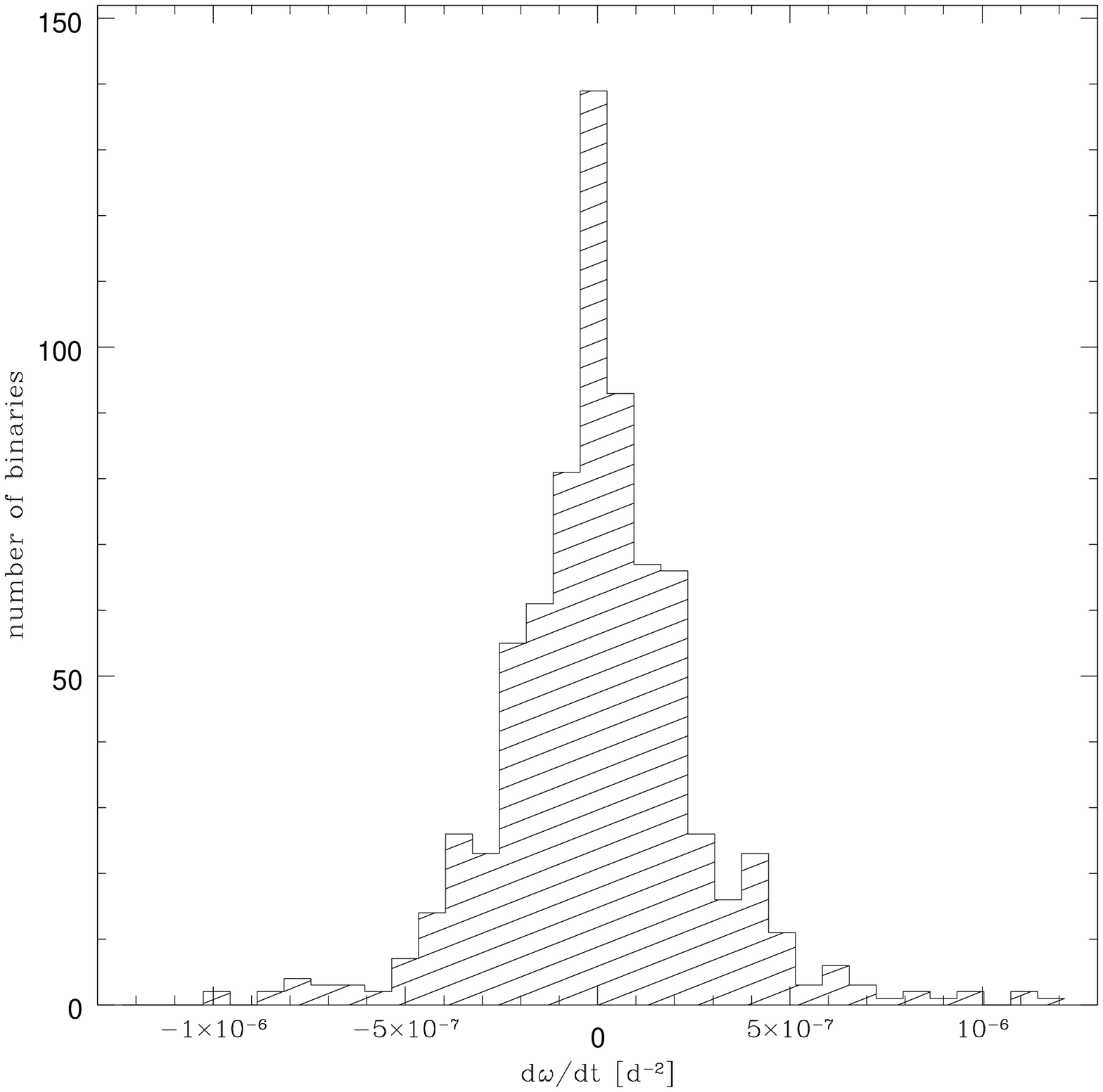}
\caption{Histogram of $\frac{d\omega}{dt}$ for 745 stars with reasonable signal to noise ratio.}
\end{figure}

For 745 stars we were able to find values of $\frac{d\omega}{dt}$
with reasonable accuracy. Distribution of this parameter is
presented in Figure 3. From these stars we chose 32 stars for which
$\frac{d\omega}{dt}$ was found with the best accuracy. This set of
stars have either high period change rate or very good signal to
noise ratio. The histogram of their $\frac{d\omega}{dt}$ is shown in
Figure 4.

\begin{figure}[ht]
\plotone{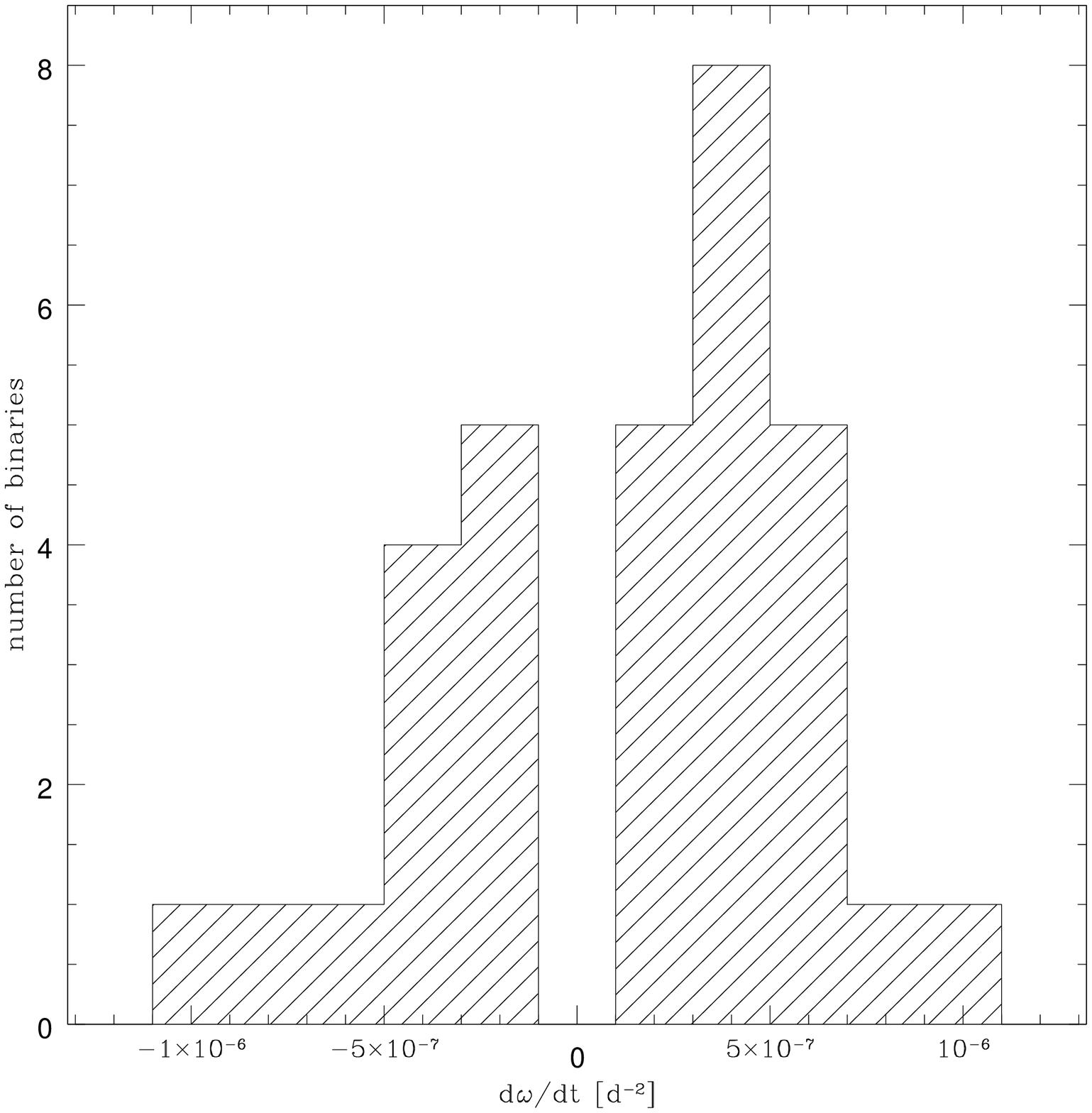}
\caption{Histogram of $\frac{d\omega}{dt}$ for stars with high and easily detectable period change rate.}
\end{figure}

As seen in Figure 3 it seems that there is no evidence that average
$\frac{d\omega}{dt}$ for short period EC and ESD binaries differs
from zero. Figure 4 shows a quite distinct asymmetry: there are more
stars with decreasing period than ones that have it increasing,
however, this sample is very small and has a little statistical
significance.

\section{Conclusion}

The ASAS data allows us to investigate a lot of very interesting
eclipsing binaries using reasonable quality data. Preliminary
results gave quite significant evidence that average
$\frac{d\omega}{dt}$ is close to zero. Method for a period change
determination used here was not perfect so we expect to give much
better results in near future as we improve our calculations.

The observations are still carried on so the amount of data is
increasing and in a few years similar calculations will give much
better results.

\section{Acknowledgements}

We thank Prof. Marcin Kubiak for his useful remarks.

\begin {references}
\reference Flannery, B. P. 1976, \apj, 305, 317
\reference Lucy, L. B, 1976, \apj, 205, 208
\reference Pojmanski, G. 2002, AcA, 52, 397
\reference Paczynski, B., Szczygiel, D., Pilecki, B., Pojmanski, G. 2006, \mnras, 368, 1311
\reference Rahunen, T. 1981, \aap, 102, 82
\reference Robertson, J. A., Eggleton, P. P. 1977, \mnras, 179, 359
\reference Schwarzenberg-Czerny, A. 1989, \mnras, 241, 153
\reference van't Veer, F. 1979, \aap, 220, 128
\reference Vilhu, O. 1982, \aap, 109, 17
\reference Wilson, R. E. 1994, \pasp, 106, 921
\end {references}
\end{document}